\documentclass[12pt]{iopart}

\usepackage[english]{babel}
\usepackage{revsymb}
\usepackage{graphicx}
\usepackage{bm}               
\usepackage{latexsym}
\usepackage{amstext,amssymb}

\begin{document}

\title[Influence of external magnetic fields on growth of alloy nanoclusters]
{Influence of external magnetic fields on growth of alloy
nanoclusters}

\author{M~Einax$^{1,2}$, S~Heinrichs$^1$, P~Maass$^2$,  A~Majhofer$^3$ and W~Dieterich$^1$}

\address{$^1$ Universit\"at Konstanz - Fachbereich Physik, D-78457 Konstanz,
  Germany}
  \address{$^2$  Technische Universit\"at Ilmenau - Institut f\"ur Physik, D-98684 Ilmenau,
  Germany}
\address{$^3$  University of Warsaw - Institute of Experimental
  Physics, Hoza 69, PL-00681 Warszawa, Poland}
  \ead{mario.einax@tu-ilmenau.de}

\begin{abstract}
  Kinetic Monte Carlo simulations are performed to study the influence of
  external magnetic fields on the growth of magnetic fcc binary alloy
  nanoclusters with perpendicular magnetic anisotropy. The underlying
  kinetic model is designed to describe essential structural and
  magnetic properties of CoPt$_3$-type clusters grown on a weakly
  interacting substrate through molecular beam epitaxy.  The results
  suggest that perpendicular magnetic anisotropy can be enhanced when
  the field is applied during growth.  For equilibrium bulk systems a
  significant shift of the onset temperature for L1$_2$ ordering is
  found, in agreement with predictions from Landau theory. Stronger field
  induced effects can be expected for magnetic fcc-alloys
  undergoing L1$_0$ ordering.
\end{abstract}

% 81.15.Aa   Theory and models of film growth
% 75.75.+a   Magnetic properties of nanostructures
% 68.55.-a   Thin film structure and morphology 
% 68.55.Ac   Nucleation and growth:  microscopic aspects
% 75.30.Gw   Magnetic anisotropy
%

\submitto{\JPCM} . %Phys: Condens. Matter

\maketitle

\section{Introduction}

It has recently been shown that CoPt$_3$ ultrathin films and
nanoclusters, grown by molecular beam epitaxy (MBE), display
perpendicular magnetic anisotropy (PMA) with potential
applications in high density magnetic storage media
\cite{Albrecht1,Albrecht2,Shapiro}. Previous simulations of
CoPt$_3$ nanocluster growth on a weakly interacting substrate
suggested that PMA originates from the combined effects of Pt
surface segregation and cluster shape
\cite{HeinrichsEPL,Heinrichs}. As suggested by experiments on
Co-Pt multilayer systems \cite{Johnson}, out-of-plane CoPt bonds
contribute to a magnetic anisotropy energy that tends to align the
Co-moments along the bond direction and thus favours PMA. When a
perpendicular magnetization is induced by a magnetic field applied
during growth, this type of a local magnetic anisotropy energy can
be expected to lead to an additional preference of out-of-plane
relative to in-plane nearest neighbour Co-Pt pairs. MBE-growth of
CoPt$_3$-clusters in a perpendicular field should therefore
improve the formation of PMA in nanoclusters.

In order to demonstrate this latter effect we perform kinetic
Monte Carlo simulations of growth within a binary alloy model,
supplemented by a magnetic anisotropy energy $H_{\rm A}$ based on
Co-Pt bond contributions. Isotropic magnetic interactions, on the
other hand, do not contribute in a significant way to any
growth-induced structural anisotropy. Their main effect in growth
simulations is a small renormalization of chemical interactions,
which can be neglected. However, both types of magnetic
interactions in Co-Pt and related fcc-alloys can lead to an
intriguing interplay between structural ordering phenomena and
magnetism in samples at equilibrium \cite{Cadeville}. We confirm
this by comparing predictions of Landau theory with equilibrium
Monte Carlo simulations.

\section{Definition of the model}

The model we consider refers to binary fcc-alloys with composition
AB$_3$. A minimal set of chemical interactions between the atoms
is chosen that is consistent with the main processes during
structure formation: atomic migration in different local
environments, surface segregation of the majority atoms and
ordering with L1$_2$-symmetry. On a semiquantitative level, these
features can be reproduced by effective chemical interactions
$V_{AA}$, $V_{AB}$ and $V_{BB}$ acting between nearest neighbour
$A$ or $B$-atoms on an fcc lattice. The linear combinations $I =
(V_{AA} + V_{BB} - 2 V_{AB})/4$, $h = V_{BB} - V_{AA}$ and $V_0 =
(V_{AA} + V_{BB})/2$ determine the bulk ordering temperature $T_0
= 1.83I/k_{\rm B}$ \cite{Binder}, the degree of surface
segregation of one atomic species and the average atomic binding
energy, respectively. Following previous work \cite{HeinrichsEPL},
we adapt our model to CoPt$_3$ (A = Co, B = Pt), where $T_0 \simeq
960\,$K, and Pt-surface segregation is strong (nearly 100 \%
\cite{Gauthier}). Compatible parameters are $I = 1$, $h \simeq 4$
and $V_0 \simeq -5$, where we have used $k_{\rm B} T_0 /1.83
\simeq 45\,$meV, corresponding to $523\,$K, as our energy unit. The
value for $V_0$ describes the average binding energy for
intermediate coordination numbers experienced by atoms within the
growth zone near the cluster surface. The substrate (111) surface
is weakly attractive with a potential $V_s = - 5$ that acts on
adatoms in the first layer. Compared to a Pt(111) surface with
three bonds of typical strength $V_0$, this amounts to about $1/3$
of the energy of a single Pt-Pt bond.

The elementary processes in our continuous time kinetic Monte
Carlo algorithm, which drive the cluster growth, are {\emph{(i)}}
codeposition of $A$- and $B$-atoms with ratio 1:3 and total flux
$F$, {\emph{(ii)}} hopping of atoms to vacant nearest neighbour
sites, and {\emph{(iii)}} direct exchange of unlike nearest
neighbour atoms; one of them is an adatom with low coordination (3
to 5) on top of a terrace and the other one a highly coordinated
atom (coordination 8 to 10) underneath. Such exchange processes
facilitate Pt segregation to the surface, which otherwise becomes
kinetically hindered through the incoming flux. In our simulations
we choose $F = 3.5$ monolayers per second. As far as possible,
other kinetic parameters are adapted from known diffusion data:
Jump rates for A and B atoms are of the form $\nu \exp[-(U +
\mbox{max}(0,\Delta E))/k_{\rm B} T]$ where $\nu \simeq 8.3 \times
10^{11}$ s$^{-1}$ is the attempt frequency, $U \simeq 5$
\cite{Bott} is the diffusion barrier, and $\Delta E$ denotes the
energy difference before and after the jump. Direct exchange
processes among pairs of unlike atoms are subjected to an
increased barrier $U + U_x$ with $U_x = 5$. For more details on
that model and on the choice of these parameters we refer to
\cite{HeinrichsEPL}.

As explained in the introduction, PMA in CoPt$_3$-nanoclusters can
be related to a structural anisotropy, expressed by
the parameter
\begin{equation}\label{parameter}
  P = (n_{\perp}^{\rm CoPt} - n_{\parallel}^{\rm CoPt} )/N ,
\end{equation}
which is defined in terms of the difference between numbers of
Co-Pt bonds out of plane, $n_\perp ^{\rm CoPt}$, and in plane,
$n_{\parallel}^{\rm CoPt}$. $N$ is the total number of atoms in
the cluster. The underlying analysis \cite{HeinrichsEPL} can in
particular account for the temperature window where PMA occurs,
which lies below the onset of L1$_2$ ordering. Furthermore, it
provides an interpretation of PMA in terms of  cluster size and
shape, and the degree of Pt-surface segregation. The associated
magnetic model assumes a local crystalline anisotropy energy
$H_{\rm A}$, that involves only Co-moments $\boldsymbol{\mu}$ as
its dominant part. Within a bond picture, each Co-atom experiences
an anisotropy energy that is a sum over Co-Pt bond contributions
\begin{equation}\label{bond}
  -A (\boldsymbol{\mu} \cdot \boldsymbol{ \delta})^2/(|\boldsymbol{ \mu}| |\boldsymbol{
  \delta}|)^2 .
\end{equation}
Here $\boldsymbol{\delta}$ is a bond vector connecting the Co-atom
with one of its nearest neighbour Pt atoms, and $A \simeq
225\,\mu$eV is a parameter deduced from the interfacial part of
the measured magnetic anisotropy of Co-Pt multilayers
\cite{Johnson}. We remark that these measurements also appear to
justify the bond model implied by (\ref{bond}), when two different
interfacial orientations are compared. For (111) interfaces the
anisotropy energy per surface area is found about twice as large
as for the (100) orientation. Considering the different angles of
bonds to the surface and the different packings, equation
(\ref{bond}) indeed can reproduce this difference with one
consistent value for $A$. Since $A > 0$, spin alignment along a
Co-Pt bond is favoured. Certainly, $A$ is small relative to the
chemical interactions $V_{AA}$, $V_{AB}$ and $V_{BB}$ as well as
$I$ and $h$. Hence, in the absence of a magnetic field it is
justified in a first approximation to neglect (\ref{bond}) in
simulating growth, but to include it {\it{a posteriori}} in order
to relate a given cluster structure to its magnetic properties.

\section{Growth in external fields}

However, when an external magnetic field is applied, we show here
that inclusion of the magnetic anisotropy energy (\ref{bond}) in
the growth process itself leads to a small but notable change in
the clusters' short range order such that PMA gets enhanced
\cite{Einax}. To demonstrate this, we apply a strong field $B_s$
perpendicular to the substrate that drives the magnetization
towards saturation. Co-moments $\boldsymbol{\mu}$ in (\ref{bond})
are then aligned along the [111] direction, which introduces an
asymmetry in the probabilities for jumps that form or break
CoPt-bonds. For in-plane bonds, (\ref{bond}) gives no
contribution, whereas the energy of an out-of-plane Co-Pt bond is
changed from $V_{AB}$ to $V^\prime_{AB} = V_{AB} - \frac{2}{3}A$
in the fcc lattice geometry.

Clearly, the expected magnetic field effects on growth are small
in view of the smallness of the magnetic anisotropy parameter $A$
in (\ref{bond}). To obtain reliable results for the change in the
structural anisotropy parameter $P$ with and without the field, we
have carried out simulations including several much larger
 parameters $A$ up to $A = 1$. Note that the energy unit
chosen (see above) implies that the physical value $A \sim 250
\mu$eV corresponds to $A = 5 \times 10^{-3}$. Results for those
different artificial $A$-values will be used in turn to extract
the physical change of $P$ by interpolation.

From this analysis we first show results for the dependence $P =
P(N,A)$ on cluster size N, see figure~1. All data sets referring to
different $A$ exhibit an approximate linearity in $N^{-1/3}$,
\begin{equation}\label{linearity}
  P(N,A) \simeq p_s(A)\, N^{-1/3}.
\end{equation}
Concerning the $N$-dependence, this confirms that the structural
anisotropy essentially is a surface effect\footnote{Fits of data
in figure~1 by straight lines and extrapolation to $N^{-1/3}
\rightarrow 0$ yield in addition a non-zero bulk contribution to
$P$ which, however, is negligibly small in comparison with
$P$-values for clusters with $N \sim 10^3$ atoms.}. The figure
includes the case $A = 0$, represented by open symbols. This data
formally corresponds to zero magnetic field, because for $B = 0$
at the temperatures considered the Co-moments $\boldsymbol{\mu}$
are oriented randomly so that (\ref{bond}) gives no contribution
to $P$.  The important observation is that $P(N,A)$ for fixed
$N$ significantly increases with $A$. Figure 2 explicitly shows that
the behaviour of the coefficient $p_s(A)$ obtained from fits
according to (\ref{linearity}), appears to be consistent with a
linear $A$-dependence. Assuming linearity even in the limit $A
\rightarrow 0$, we  obtain $(p_s(A) - p_s(0))/p_s(0) \simeq 3
\cdot 10^{-2}$ for the physical value $A = 5 \times 10^{-3}$.

\begin{figure}
\begin{center} \includegraphics{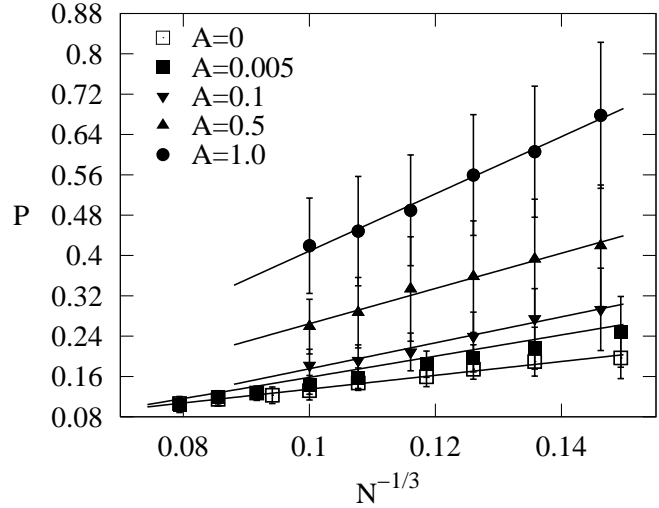} \end{center}
\caption{Simulated structural anisotropy
  parameter $P$ depending on cluster size $N$ in the presence of the
   saturation field $B_s$ at temperature $k_B T = 1.2$ $(T \simeq 630$K).
  Averages are performed over 20 clusters. Full lines display
  linearity in $N^{-1/3}$. Open symbols are equivalent to the case of zero magnetic field.}
 \label{f.1}
\end{figure}

This result immediately translates to the magnetic
field-dependence of the structural part of the total magnetic
anisotropy energy $E_s$, which is proportional to $N P$. Using
(\ref{linearity}), we  can write
\begin{equation}\label{anisoenergy}
  E_s = K_s(B) N^{2/3},
\end{equation}
again with the relative change $ (K_s(B_s) - K_s(0))/K_s(0)
\simeq 3 \cdot 10^{-2}$. Thus, our main result  is that the
surface anisotropy constant $K_s$ in (\ref{anisoenergy}) increases
by about 3 $\%$ when a perpendicular field of the strength of the
saturation field $B_s$ is switched on. Near $T \simeq 1$ ($T \simeq 523
K$), where the anisotropy develops its maximum \cite{Einax}, we obtain 
analogous results (not shown).

\begin{figure}
\begin{center}\includegraphics{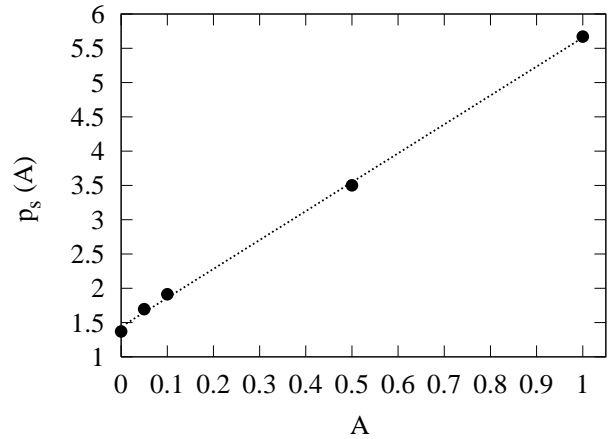} \end{center}
\caption{Coefficient $p_s(A)$ determining the structural
anisotropy versus the magnetic anisotropy constant $A$ at $T =
1.2$. Data points result from the slopes of figure~1 including 
the case $A=0.05$. Dotted line: linear fit.} \label{f.2}
\end{figure}

\section{Interplay of structural and magnetic order in the bulk}

So far we have shown that in the presence of a strong
perpendicular magnetic field the anisotropy energy $H_{\rm A}$
notably affects the frozen-in structure of growing nanoclusters
and thus improves PMA. Based on Landau theory, we now study
effects of external magnetic fields on the equilibrium structure
of bulk systems or homogeneous films. Effects of this type are
known in principle but have been explored only for few specific
materials \cite{Chika}--\cite{Dang}. For CoPt$_3$, quantitative
calculations have been performed in the past within the cluster
variation method in the tetrahedron approximation \cite{Sanchez}.
Landau theory, which we use here allows us to incorporate the
magnetic anisotropy in a simple manner and helps clarifying the
interrelation between different effects.

For the free energy $f$ per Co-atom in CoPt$_3$ in the presence of an
external field $\mathbf{ B}$ we propose the form
\begin{equation}\label{f}
f = f_S + f_M + f_{\rm int} - \mathbf{ M} \cdot \mathbf{ B } ,
\end{equation}
where $\mathbf{ M }= (M_1, M_2, M_3)$ is the average magnetic moment
per CoPt$_3$-unit. The first term in (\ref{f}) denotes the structural
free energy that describes L1$_2$-ordering without magnetic
contributions \cite{Lai},
\begin{equation}\label{fs}
f_S = \sum^3_{\alpha = 1}\left( \frac{r (T)}{2}\psi_\alpha^2 +
\frac{v}{4}\psi^4_\alpha \right) + \frac{u}{4}\left(
\sum^3_{\alpha = 1}\psi^2_\alpha\right)^2 + w \psi_1\psi_2\psi_3.
\end{equation}
The structural order parameter components $\psi_\alpha$, $\alpha =
1,2,3$, describe amplitudes for layering along one of the cubic axes
$\alpha$ of the fcc-lattice such that CoPt-layers and Pt-layers
alternate. Superposition of these layer structures in all three
directions yields the L1$_2$-structure. As usual, we set $r(T) \simeq
r_0 (T - T_{\rm sp})$; $r_0 > 0$, $u > 0$, $0 < v < u$ and $w > 0$,
where $T_{\rm sp}$ denotes the spinodal temperature. The associated
ordering temperature $T_0$ is determined by $r(T_0) = (2/9)w^2/(3u +
v)$.

The second term in (\ref{f}) is the isotropic magnetic free
energy, with the behaviour
\begin{equation}\label{mag}
f_M(\mathbf{ M}) \sim \frac{b(T)}{2}\mathbf{ M}^2,
\end{equation}
as $\mathbf{M} \rightarrow 0$, where $b(T) \simeq b_0 (T-T_C)$ and
$T_C$ denotes the Curie-temperature of a disordered alloy. Finally the
third term in (\ref{f}), $f_{\rm int}$ describes the coupling between
structural and magnetic order. Retaining only the lowest-order terms
allowed by symmetry, we write
\begin{equation}\label{lowest}
  f_{\rm int} = \frac{c_1}{2}\left( \sum^3_{\alpha=1} \psi^2_\alpha
  \right) \mathbf{ M}^2 + \frac{c_2}{2}\sum^3_{\alpha = 1}
  \psi^2_\alpha M^2_\alpha.
\end{equation}
For later purposes we define the Landau coefficients such that
$\psi_\alpha$ becomes dimensionless, with $|\psi_\alpha| = \psi_{\rm
  max} = 1$ for perfect structural order. Expressions
(\ref{f})~--~(\ref{lowest}) imply far-reaching interdependencies
between structural and magnetic properties. Combination with simple
mean-field arguments allows us to make order-of-magnitude
predictions which can be tested experimentally.

First, collecting terms in (\ref{f}) which are quadratic in
$\psi_\alpha$, we find that in a state magnetized parallel to the
[111] direction, both the spinodal and the transition temperature
for L1$_2$-ordering are shifted by an amount $T_{\rm sp}(M) -
T_{\rm sp} \simeq T_0(M) - T_0 = \Delta T_0 (M)$ with
\begin{equation}\label{down}
  \Delta T_0 (M) = - (3c_1 + c_2)M^2 / 3 r_0.
\end{equation}
To determine the $c$-coefficients in (\ref{down}) we note that
the same combination $3c_1 + c_2 $ enters the difference in Curie
temperatures between disordered and ordered samples,
\begin{equation}\label{quadratic}
 \Delta T_C (\psi_{\rm max}) = T_C (\psi_{\rm max}) - T_C = - (3 c_1 +
 c_2)/b_0 .
\end{equation}
This relation is immediately obtained from (\ref{f}) by requiring the
terms quadratic in $\mathbf{M}$ to vanish. For the ordered samples we
have set $|\psi_\alpha| \simeq \psi_{\rm max}$, because the
discontinuity $\psi_0$ in $\psi_\alpha$ at $T_0$ is large and Curie
temperatures are considerably lower than $T_0 \simeq 960\,$K. Within
the same consideration the inverse magnetic susceptibility $\chi^{-1}$
develops a discontinuity at the order transition at $T_0$ of magnitude
$(3c_1 + c_2)\psi_0^2$.  Experimentally, $T_C(\psi_{\rm max}) \simeq
400\,$K, while $\Delta T_C(\psi_{\rm max}) \gtrsim 150\,$K
\cite{Sanchez}. The reason for an increased Curie temperature in
quenched, disordered samples is the increased number of Co-Co direct
exchange couplings relative to the ordered state. This temperature
shift already allows a rough estimate of (\ref{down}). Suppose, for
example, the magnetic field is strong enough to drive the magnetic
moment $M$ to values near its saturation value $M_s$. Using mean field
arguments, we expect that both $r_0/k_{\rm B}$ and $b_0 M_s^2/k_{\rm
  B}$ are of order unity. Therefore the two expressions
(\ref{quadratic}) and (\ref{down}) should be of similar magnitude.
In other words, for very strong magnetic fields $\mathbf{ B}
\parallel [111]$ equation (\ref{f}) predicts a lowering of $T_0$ by
about $\Delta T_{0,\rm max} \simeq 10^2\,$K in CoPt$_3$.

A further effect is worth mentioning. When the orientation of the
external field is changed from [111] to [001] and $c_2 < 0$, the
disordered phase develops an instability towards formation of a
layered structure with $\psi_1 = \psi_2 = 0$ and $\psi_3 \neq 0$,
i.e. with $L1_0$ symmetry. The corresponding transition point is
increased relative to the spinodal temperature for [111] field
orientation, $T_{\rm sp}(M)$, by an amount $(2/3)|c_2|$, but will
exceed $T_0(M)$ only if $|c_2|$ is sufficiently large. Otherwise
this layered structure cannot form because the conventional
ordering transition at $T_0(M)$ will set in before. In CoPt$_3$
anisotropic magnetic interactions determining $c_2$ are weak (see
below) so that this effect is unlikely to be observable.

At this point let us attempt to explicitly relate the parameters $c_1$
and $c_2$ to a specific magnetic model. For that purpose we adopt a
phenomenological spin-Hamiltonian $H_M = H_{\rm ex} + H_{\rm A}$
pertaining to CoPt$_3$ alloys. As its main contribution we assume
classical Heisenberg-type isotropic exchange interactions $H_{\rm ex}$
within nearest-neighbour Co-Co and Co-Pt pairs. This simplified form
seems sufficient as we are dealing only with the energetics of
homogeneously magnetized states. The respective magnetic energies for
Co-Co and Co-Pt pairs with parallel spins are $-JM^2_s$ and $-J'
M^2_s$, respectively\footnote{Pt-moments have been shown to depend on
the actual atomic environment \cite{Sanchez} but this effect is
neglected here.}. In addition, an anisotropy 
$H_{\rm A}$ connected with Co-Pt bonds, see (\ref{bond}), is included.  
Within first order
perturbation theory the change in free energy due to $H_M$ is given by
an average of $H_M$ over an unperturbed ensemble specified by the
$\psi_\alpha$. Mean-field calculations then yield
\begin{equation}\label{mean}
c_1 = J - 2 J^\prime, \quad c_2 = - 2A/M_s^2.
\end{equation}

In order to test the validity of the Landau approach based on the
coupling (\ref{lowest}) between structural and magnetic order and its
prediction (\ref{down}), we first focus on the anisotropy energy 
$H_{\textrm{A}}$ (ignoring $H_{\textrm{ex}}$) and
perform Monte Carlo simulations of the lattice model described before,
but now under equilibrium conditions. The simulation box contains
43200 atoms and is subjected to periodic boundary conditions in all
three directions. Equilibrated configurations are generated by
allowing any pair of unlike atoms to exchange positions.
\begin{figure}
\begin{center}\includegraphics{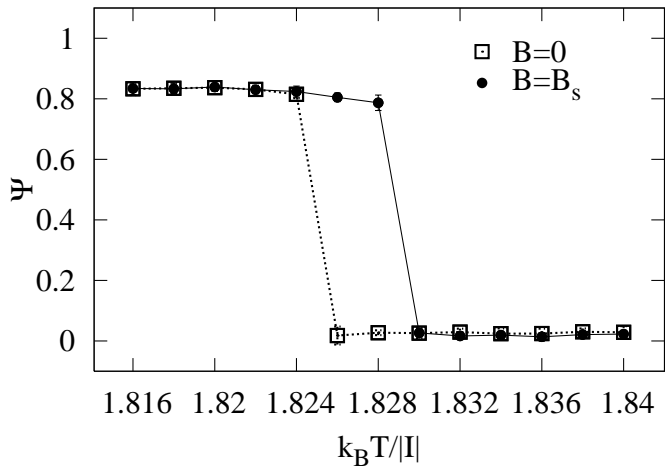}\end{center}
\caption{Simulated L1$_2$ order parameter $\Psi (T)$ for a
  bulk system at equilibrium as a function of temperature. Averages
  were performed over 6 realizations. Among the magnetic interactions
  only the anisotropy $H_A$ has been included in the simulations.  The
  upward shift of the ordering temperature $T_0(B) - T_0 \simeq 2\,$K
  is clearly recognizable. } \label{f.3}
\end{figure}
In figure~3 the structural order parameter is plotted versus
temperature, and indeed shows an increased transition temperature in
the fully magnetized state in comparison to the zero field case. We
have again used $A \simeq 225\,\mu$eV, see above, which gives an
increase of $T_0$ by $2\,$K. This increase favourably agrees with
(\ref{down}) when we set $c_1 = 0$, see (\ref{mean}), and $r_0/k_{\rm
  B} \simeq 1$. These results show that with realistic parameters for
the microscopic anisotropy energy a measurable shift in the ordering
temperature is observed. The physical reason for the increase of $T_0$
in figure~3 is that through the presence of $H_{\rm A}$ the six
nearest neighbour sites of a Co-atom with bond vectors not orthogonal
to [111] become preferentially occupied by Pt-atoms during the
equilibration process, thus favouring the ordered structure. Basically
this is the same mechanism that yields an increased structural
anisotropy parameter $P$ discussed before. Figure~3, therefore,
provides additional evidence, that $H_{\rm A}$ affects the ordering
transition.

When exchange interactions are included, equation (\ref{down})
predicts a much larger downward shift in $T_0$, which we can estimate
now in a more quantitative manner. From \cite{Sanchez} and known Co-
and Pt-moments we obtain $JM_s^2 \simeq 140\,$K and $J'M_s^2 \simeq
25\,$K, while $A$ is much smaller. Thus $(3c_1 +c_2)M_s^2/3 \simeq
90\,$K, in agreement with the order of magnitude estimate of
(\ref{down}) given before, which was based on the comparison with
measured Curie temperatures.

 Again, the corresponding change in $T_0$ according to
(\ref{down}) can be compared with predictions from numerical
computation. However, within a lattice-model with nearest-neighbour
chemical and isotropic exchange interactions, extra simulations
for this problem actually are not needed when we note that in a
fully magnetized sample the effect of $H_{\rm ex}$ is simply
equivalent to modified chemical interactions
\begin{equation}\label{chemical}
 V'_{AA} = V_{AA} - JM^2_s; \quad V'_{BB}= V_{BB}; \quad V'_{AB} =
  V_{AB} - J' M^2_s .
\end{equation}
Since the ordering transition temperature satisfies $k_{\rm B} T_0 \simeq
1.83(V_{AA} + V_{BB} - 2 V_{AB})/4$, we find, using (\ref{chemical}),
$k_{\rm B}\Delta T_0 (M_s) \simeq -1.83 (J-2J')M_s^2/4$. Because of
(\ref{mean}) and $|c_2| \ll c_1$ this is consistent with (\ref{down})
and establishes a semi-quantitative estimate $\Delta T_0(M_s) \simeq
-41\,$K. Clearly, since the exchange terms $JM^2_s$ and $J^\prime M^2_s$
in (\ref{chemical}) are relatively small, we could neglect them in our
previous discussion of structural anisotropy.

Experimentally, for CoPt$_3$ it is probably very difficult to
induce a sufficiently strong magnetization near $T_0$ such that
the shift $\Delta T_0 \propto M^2$ becomes measurable. The
situation is more favourable for L1$_0$-ordering of
Co$_x$Pt$_{1-x}$ alloys near $x = 0.5$, where the Curie
temperature is closer to the ordering temperature $T_0$ and even
intersects with $T_0$ on the Co-rich side in the phase diagram. A
similar situation occurs for Ni-rich Ni$_x$Pt$_{x-1}$ alloys
\cite{Cadeville}. For L$1_0$-ordering the structural part of the
free energy $f_S$ is again based on (\ref{fs}), however, with a
change in sign in $w$ at $x = 1/2$, $v$ sufficiently negative and
additional stabilizing higher order terms \cite{Tano}. Elastic
energy contributions due to the tetragonal distortion in the
L1$_0$ phase are thereby neglected. Coupling between structural
order parameters and the magnetization can again be represented by
equation (\ref{lowest}). Considering magnetic fields $\textbf{B}
\parallel [001]$ we now obtain $\Delta T_{\rm sp}(M) \simeq \Delta
T_0(M) = - (c_1 + c_2)M^2/r_0$ and $\Delta T_C (\psi_{\rm max}) =
- (c_1 + c_2)/b_0$ instead of (\ref{down}) and (\ref{quadratic}),
with a measured value of about $10^2\,$K for the latter
temperature shift. Using similar arguments as before, both
temperature shifts $\Delta T_{\rm sp} (M_s)$ and $\Delta T_c
(\psi_{\rm max})$ are expected to be of the same order.

\section{Conclusions and outlook}
Effects of external magnetic fields on both the
far-from-equilibrium MBE growth of magnetic fcc alloys and their
bulk structural phase behaviour at equilibrium have been studied.
The statistical model we employed pertains to binary alloys of the
CoPt$_3$ type and L1$_2$-ordering. KMC simulations are
supplemented by equilibrium considerations based on Landau theory.
Using realistic parameters for CoPt$_3$ our main aim was to
explore the influence of the local crystalline magnetic anisotropy
energy, modelled by a bond Hamiltonian $H_A$, on the structural
properties of alloy nanoclusters, when a strong magnetic field is
applied during growth. Such an effect, although small in CoPt$_3$,
appears to be general and indeed leads to a notable enhancement of
the clusters structural anisotropy and the associated PMA.
Experimentally, in order to induce a large magnetization, the
substrate temperature should fall below the Curie temperature of
disordered samples but at the same time remain in the known
temperature window where PMA occurs.

More favorable conditions for studying these effects
experimentally should exist for alloys like CoPt or FePt
\cite{Massalski} undergoing L1$_0$-ordering, if clusters could be
grown along the c-axis. As known from thin film measurements
\cite{Iwata}, the magnetic anisotropy in those alloys is a bulk
property connected with alternating Co(Fe)- and Pt-rich layers caused by 
the L1$_0$ structural order. Hence, under growth conditions where 
L1$_0$-ordering is suppressed, one can expect a substantially
larger field-induced PMA.

\ack We thank M. Albrecht and G. Schatz for fruitful discussions.
This work has been supported by the Deutsche
Forschungsgemeinschaft DFG (SFB 513).


\section*{References}
\begin{thebibliography}{999}

\bibitem{Albrecht1}  Albrecht M, Maier A, Treubel F, Maret M, Poinsot P  and Schatz
G 2001 {\it{Europhys. Lett.}} {\bf 56} 884

\bibitem{Albrecht2}  Albrecht M, Maret M, Maier A, Treubel F, Riedlinger B, Mazur U, Schatz G.
and Anders S 2002
  {\it{J. Appl. Phys.}} {\bf 91} 8153

\bibitem{Shapiro}  Shapiro L, Rooney P W, Tran M Q, Hellman F, Ring K M, Kavanagh K L, Rellinghaus B
 and Weller D 1999
  {\it{Phys. Rev. B}} {\bf 60} 12826

\bibitem{HeinrichsEPL} Heinrichs S, Dieterich W and Maass P 2006
  {\it{Europhys. Lett.}} {\bf{75}} 167

\bibitem{Heinrichs} Heinrichs S, Dieterich W and Maass P 2006
  {Epitaxial growth of binary alloy nanostructures} \textit{Preprint} 
  cond-mat/0607284

\bibitem{Johnson} Johnson M T, Bloemen P J H, den Broeder F J A and de Vries J J 1996
  {\it{Rep. Prog. Phys.}} {\bf 59} 1409

\bibitem{Cadeville}  Cadeville M C and Mor\'an-Lop\'ez J L 1987
  {\it{Phys. Rep.}} {\bf 153} 331

\bibitem{Binder}  Binder K 1980
  {\it{Phys. Rev. Lett.}} {\bf 45} 811

\bibitem{Gauthier}Gauthier Y, Baudoing-Savois R, Bugnard J M, Bardi U and Atrei
  A 1992  {\it{Surf. Sci.}} {\bf 276} 1

\bibitem{Bott}Bott M, Hohage M, Morgenstern M, Michely T and Comsa
  G 1996
  {\it{Phys. Rev. Lett.}} {\bf 76} 1304

\bibitem{Einax} Einax M, Heinrichs S,
  Maass P, Majhofer A, and Dieterich W
   {\it{Proc. E-MRS 2006
  Spring Meeting (Nice), Material Science and Engineering:
  C,}} in press

\bibitem{Chika} Chikazumi S and Graham C D 1999
  {\it Physics of Ferromagnetism} 2nd edition
    (Oxford: Oxford University Press)

\bibitem{Dang}  Dang M Z and Rancourt D G 1996
  {\it{Phys. Rev. B}} {\bf 53} 2291

\bibitem{Sanchez}
  Sanchez J M, Mor\'an-L\'opez J L, Leroux C and Cadeville M C
  1989   {\it{J. Phys. C: Condens. Matter}} {\bf 1} 491

\bibitem{Lai}  Lai Z W 1990
  {\it{Phys. Rev. B}} {\bf 41} 9239

\bibitem{Tano} Tano\u{g}lu G B, Braun R J, Cahn J W and McFadden G
  B 2003 {\it{Interfaces and free boundaries}} {\bf{5}} 275
 
\bibitem{Massalski}  Massalski T, 1996 {\it{Binary alloy phase
  diagrams}} vol~2 (Ohio: ASM International)

\bibitem{Iwata} Iwata S, Yamashita S and Tsunashima S 1997
  {\it{IEEE Transactions on Magnetics}},  {\bf{33}}  3670
\end{thebibliography}
\end{document}